\documentclass[12pt]{article}

\usepackage{scicite}
\usepackage{hyperref}

\usepackage{ragged2e}

\justifying
\usepackage{times}

\usepackage{verbatim}
\usepackage{chemformula} 
\usepackage[T1]{fontenc} 
\usepackage[version=3]{mhchem} 
\usepackage[T1]{fontenc}       
\usepackage{graphicx} 
\usepackage{subcaption} 
\usepackage[separate-uncertainty = true, multi-part-units = single, list-units = single, range-units = single]{siunitx} % units
\DeclareSIUnit\Molar{\textsc{m}}
\DeclareSIUnit\rpm{rpm}
\DeclareSIUnit\ppm{ppm}
\DeclareSIUnit\kbt{k_BT}
\usepackage{geometry}

\usepackage[noabbrev,nameinlink]{cleveref}
\usepackage{multirow}
\usepackage{makecell}
\usepackage{colortbl}

\usepackage{stackengine}
\usepackage{svg}

\newenvironment{sciabstract}{%
	\begin{quote} \bf}
	{\end{quote}}

\title{Optimal self-assembly pathways towards colloidal lattices with tunable flexibility} 

\author
{Yogesh Shelke,$^{1}$ Daniel J.G. Pearce,$^{2}$ Daniela J. Kraft$^{1}$$^\ast$\\
	\\
	\normalsize{$^{1}$Soft Matter Physics, Huygens-Kamerlingh Onnes Laboratory, Leiden University, The Netherlands}\\
	\normalsize{PO Box 9504, 2300 RA Leiden, The Netherlands.}\\
	\normalsize{$^{2}$Department of Theoretical Physics, \'Ecole de Physique, University of Geneva, Switzerland}\\
	\normalsize{24 Quai Ernest-Ansermet, 1211 Geneva, Switzerland.}\\
	\\
	\normalsize{$^\ast$To whom correspondence should be addressed; E-mail: kraft@physics.leidenuniv.nl}
}

\date{}

\begin{document}

	\baselineskip24pt

	\maketitle

	\begin{sciabstract}
		Flexibility governs the many properties of materials and is crucial for the function of proteins and biopolymers. However, how the self-assembly of flexibly bonded particles can lead to larger structures with global reconfigurability is unexplored. We here use a binary colloidal model system equipped with flexible DNA-based bonds to study how regular structures with tunable flexibility can be created through self-assembly. We find that the reconfigurability during lattice growth leads to lattices with square symmetry which are inherently mechanically unstable and hence thermally floppy. By considering the role of size ratio, number ratio, and directionality induced by particle shape, we identify the optimal pathways that maximize the yield and flexibility of these square lattices using a combination of experiments, analytical calculations, and simulations. Our study highlights the crucial role of reconfigurability in systems that are governed by enthalpic and entropic principles, from synthetic to biological, and might be useful for creating materials with novel or reconfigurable properties.  
		\\
	\end{sciabstract}

	\section*{INTRODUCTION}
	
	Bond flexibility is crucial in materials and structures at all length scales. At the atomic and molecular scale, small rearrangements of the atoms in a material are key to heat and sound transport, and influence the mechanical, thermal and electronic properties of materials~\cite{williams2007off, wu2023stimuli,wang2019flexible,wang2015crystallization,krishna2016mechanically,smallenburg2013liquids}. In macroscopic materials, hinging between the constituent units has recently been employed to create a new generation of materials with designer properties, so-called mechanical metamaterials.\cite{bertoldi_flexible_2017} At the intermediate micrometerlength scale of colloidal structures, rearranging bonds have long been elusive. Although theoretical and numerical studies typically assume rearrangements being possible after bonding, in experiments this is often not the case. Most colloidal bonds do not allow rearrangements due to surface inhomogeneities which trap their configuration in metastable, rigid states. This restricts an experimental comparison with theoretical predictions for example about the effect of flexibility on mechanical properties\cite{kohlstedt_self-assembly_2013} or on the phase behavior\cite{smallenburg2013liquids, smallenburg_erasing_2014} and creating functional microscale structures that do work through conformational changes.\cite{brandenbourger_limit_2022}

	Recently, two strategies have emerged to experimentally obtain rearranging bonds between colloidal particles. In the first approach, the temperature range and interaction strength is precisely controlled such that colloidal particles functionalized with DNA linkers covalently bound to the particles' surface can rearrange.\cite{wang2015crystallization, wang_synthetic_2015} This is made possible by creating weak, transient DNA bonds. The power of rearrangements have been demonstrated by showing that these particles can self-assemble into crystalline lattices that are otherwise not achievable by DNA-based interactions.\cite{wang2015crystallization, wang_synthetic_2015} An additional benefit is that the lattices become rigid and preserve their conformation after just a slight cooling.  
	
	The second strategy employs DNA linkers that are mobile on the colloid surface.\cite{van2013solid,mcmullen2018freely,rinaldin2019colloid} The linkers can be either placed on fluid droplets \cite{mcmullen2018freely} or inside lipid bilayers supported by solid colloidal particles to be laterally mobile over a wide temperature range.\cite{van2013solid, rinaldin2019colloid} Although mobile, the bond strength for surface mobile DNA linkers can be set by designing the DNA sequence to obtain any desired bond strength including strong, effectively irreversible bonds, which in addition can further be tuned by temperature \textit{in situ}. This strategy has been employed to create a variety of flexible structures ranging from flexible chains to rings and colloidal molecules - with and without limited angular range - up to foldable structures and floppy square lattices.\cite{chakraborty2017colloidal,verweij2020flexibility,verweij2021conformations,chakraborty2022self,verweij2023brownian,shelke2023flexible,mcmullen2022self} The latter consisted of square arrangements of colloidal particles obtained by assembling the particles one-by-one using optical tweezers.\cite{melio2024soft} Excitingly, the multivalent DNA-based bond was shown to act as a microscopic spring, making this a model system for thermal spring networks and allowing the extraction of the floppy and rigid normal modes.\cite{melio2024soft}
	However, the manual assembly strategy like the one adopted here is limited to small structures, with the largest structure achieved so far being a 7x7 lattice.
	
	A more powerful strategy for creating larger structures is self-assembly.\cite{gartner2024design} Self-assembly exploits the Brownian motion of the constituents and is driven by a lowering of the free energy. Bond flexibility has been leveraged for the self-assembly of flexible colloidal clusters\cite{chakraborty2022self,shelke2023flexible} and crystalline lattices.\cite{wang2015crystallization, wang_synthetic_2015, oh_reconfigurable_2020} Rearrangements played a crucial role: for the colloidal clusters lateral motion after binding enabled reaching the maximum geometrically possible cluster size and thus high yields.\cite{chakraborty2022self, shelke2023flexible} For crystalline lattices rearrangements were important for obtaining highly ordered structures.\cite{wang2015crystallization, wang_synthetic_2015, oh_reconfigurable_2020} However, because the latter were created from surface-bound DNA linkers, their flexibility only persists for transient bonds and can therefore not be maintained outside a narrow temperature range. This makes it challenging to use as a model system for understanding floppy modes in thermal systems and creating functional colloidal structures that, for example, execute work by changing their conformation. 
	
	Employing self-assembly in colloids with fully flexible bonds therefore has great potential to combine the advantages of both systems: on the one hand self-assembly may allow the creation of many large structures at the same time and control over their crystal symmetry through the design of the interactions, while at the same time maintain the bond flexibility and introduce floppiness in the assembled structures, at least when the bond network is appropriately designed. However, the impact of rearranging bonds on the kinetics of lattice formation and  how to guide self-assembly into floppy structures are as of yet unanswered questions. 
	
	Here, we experimentally study the nucleation and growth dynamics of floppy crystalline structures from a binary mixture of colloidal particles with flexible bonds with number and size ratios close to one. We investigate the influence of bond flexibility for equal sized spherical particles and find that their ability to rearrange enhances the formation of two-dimensional square lattices. Utilizing \textit{in situ} observations with confocal microscopy, we map out their evolution to gain a deeper understanding of the pathways leading to the formation of these flexible colloidal lattices. We leverage particle size ratio and shape to guide the self-assembly towards square lattices with high fidelity using simulations and analytical results. Our findings may serve as a tool to design innovative flexible materials with tailored properties that may be employed as sensors, actuators and in smart devices and machines. 
	
	\section{RESULTS}

	\subsection{Self-assembly of equal sized spheres}
\begin{figure}[t!]
		\centering
		\includegraphics [width=0.99\textwidth]{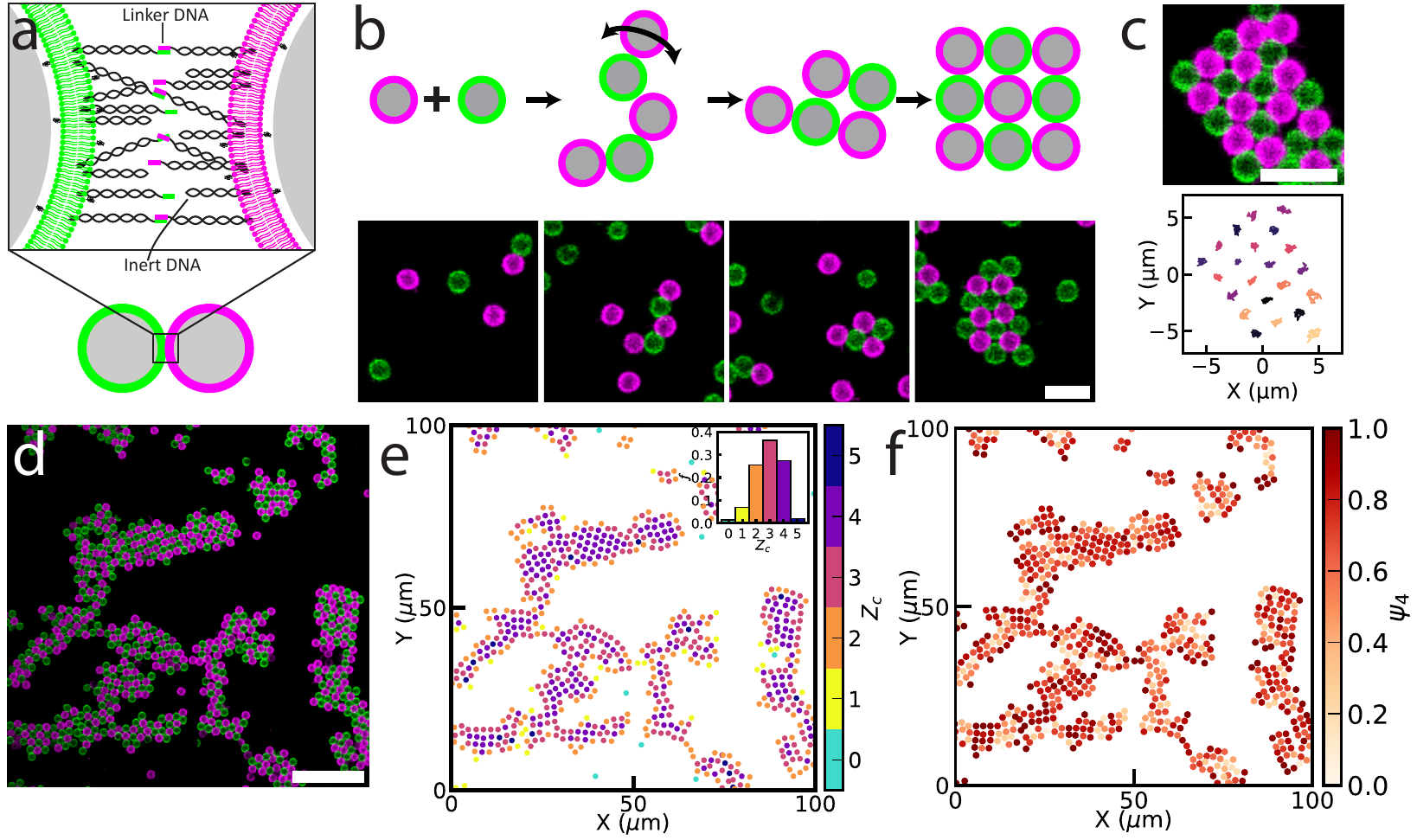}
		\caption{\small \textbf{Self-assembly of flexible colloidal lattices.} (a) Schematic of a pair of flexibly bonded colloidal spheres functionalized with a lipid bilayer and surface mobile inert and complementary linker DNA (green and magenta). (b) (top row) Schematic showing a possible reconfiguration which during during cluster growth can lead to ordered, floppy structures with a square symmetry(top row). This is mirrored by experimental observations on a binary mixture of equal sized spheres with 2.06$\pm$0.05 $\mu$m  diameter (bottom row); complementary DNA functionalization is indicated in magenta and green. (c) The corresponding confocal microscope snapshot of an exemplary lattice and trajectories of individual particles within this lattice for 100 sec from which rotational and translational motions have been subtracted demonstrating the flexibility of the lattice, see also SI Movie 3. (d) Confocal microscopy snapshots of final self-assembled lattice with equal sized spheres, see also SI Figure S1. The same snapshot is shown colored by (e) coordination number, $Z_c$, and (f) fourth order parameter, $\psi_4$.}
		\label{fig:equal_size}
	\end{figure}

	To study the influence of rearranging bonds on nucleation and growth of floppy square lattices, we experimentally coat colloidal beads with a fluid lipid membrane into which DNA linkers are embedded.\cite{van2013solid,chakraborty2017colloidal,rinaldin2019colloid}  This embedding allows the linkers to laterally move over the surface of the bead which permits rearrangements after bonding. Each bead is functionalized with one of two complimentary DNA linkers, such that when they come in close contact the DNA linkers bind and the beads become connected, see Fig.~\ref{fig:equal_size}a (See Materials and Methods, and SI for experimental details).  
	Since the DNA linkers are able to freely move within the fluid membrane enveloping a bead, the angle between bound beads is able to freely change by thermal fluctuations, giving the bonds the required flexibility, see Fig.~\ref{fig:equal_size}b (top row). 
	
	We first study the self-assembly of a system of two types colloidal spheres with equal size (diameter d=2.06$\pm$0.05$\mu$m) and concentration corresponding to a 2D area fraction $\phi$$\approx$$20\%$, but complementary DNA linker type (6bp sticky end); see See Materials and Methods for experimental details. 
	After mixing the two species in a temperature controlled setup, we observe their assembly in time by confocal microscopy, where different DNA functionalizations can be identified by labelling with different dyes. Quickly after the start of the experiment, the beads assemble into small clusters which continue to grow through addition of single particles and cluster-cluster merging, see Fig.~\ref{fig:equal_size}b (bottom row) and SI movies 1 and 2. Interestingly, a high fraction of these structures resembles small A-B type square lattices, in which the oppositely functionalized beads form a chessboard like pattern, see Fig.~\ref{fig:equal_size}b-d as well as SI movie 2 and SI Figure S1. Larger networks with square symmetry show relative motion of the particles due to the flexibility of the bonds and the arrangements of the particles. The resulting bond network is not mechanically stable with on average four neighbors, and therefore shows floppiness, see  Fig.~\ref{fig:equal_size}c (bottom), where traces of the particle positions in time are shown, as well as SI Movie 3.
	
	We analyse the clusters that have formed after $\approx$2-16 hours by calculating the coordination number, $Z_c$, defined as the number of beads bound to a given sphere. For equal sized spheres, $Z_c$ takes a maximum value of $Z_c=6$ for close packed spheres and $Z_c=4$ for a square lattice in the bulk; at the boundary of a lattice the coordination number is halved. For a system of clusters each consisting of a lattice with coordination number $Z_c^{\rm{lat}}$ the mean coordination number is approximated by:
	\begin{equation}
		\langle Z_c \rangle \approx Z_c^{\rm{lat}}\left[1 - \frac{\langle\gamma\rangle}{2}\right].
	\end{equation}
	Where we have introduced the boundary ratio $\gamma\in[0,1]$ which represents the average proportion of spheres on the boundary of a cluster. This depends on the configurations of the clusters as well as their size; see SI for further discussion. In our experiments we observe a range of coordination numbers peaked around $Z_c=3$ and an estimated boundary ratio with a value of $\gamma \approx 0.6$ which is compatible with square lattices ($Z_c^{\rm{lat}}=4$), see Fig.~\ref{fig:equal_size}e. The presence of values smaller than four stems from the small size of the clusters where many particles are located at the boundary and hence have less than four neighbors. 
	
	To capture the degree to which the arrangement of particles have rotational symmetry of order 4, we introduce a square order parameter
	\begin{equation}
		\psi_4 =\langle\tfrac{1}{n_i}|\sum_{j}^{n_i} \exp (4i\theta_{ij})|\rangle_i.
	\end{equation}
	Where the sum runs over the $n_i$ particles bound to particle $i$ and $\theta_{ij}$ is the angle of the bond connecting them; angular brackets $\langle\rangle_i$ denotes an average over all particles $i$ with $n_i>1$. Our experiments show large regions where $\psi_4>0.5$ indicating a high degree of square order symmetry and a A-B lattice like arrangement, see Fig.~\ref{fig:equal_size}f. 
	
	In a square lattice, each particle has on-average four neighbors. This is lower than the geometrically possible maximum of six and the for entropic reasons maximal achievable number of five bound particles which we previously identified.\cite{chakraborty2022self} It is also not the same as any of the lattice planes of the CsCl lattice previously found in 3D for equal sized colloids with complementary surface bound DNA linkers \cite{wang2015crystallization, wang_synthetic_2015, oh_reconfigurable_2020} and colloids with opposite charges\cite{leunissen_ionic_2005}. The observation of a square symmetry can furthermore not be fully explained by thermodynamic arguments for which an A-B lattice is expected in 2D at equal particle concentrations, because the bonds in our system are essentially irreversible. Therefore, particles cannot unbind to optimize their position in the lattice and find the thermodynamically expected state. However, while maintaining the bond they are still able to move with respect to each other to increase the number of bonds and lower their free energy. For example, a chain of at least four particles can rearrange and bind to form a small square, see Figure \ref{fig:equal_size}b. We therefore hypothesize that the formation of the square lattice stems from a combination of free energy minimization at the one hand and a competition between the timescales for adding another particle and reconfiguration of the existing structure on the other. This implies that the self-assembly pathways towards square lattices play an important role.
	
	\clearpage

	\subsection{Self-assembly pathways of flexible colloidal lattices for spheres of equal size and number ratio}
	
\begin{figure}[b!]
		\centering
		\includegraphics [width=0.99\textwidth]{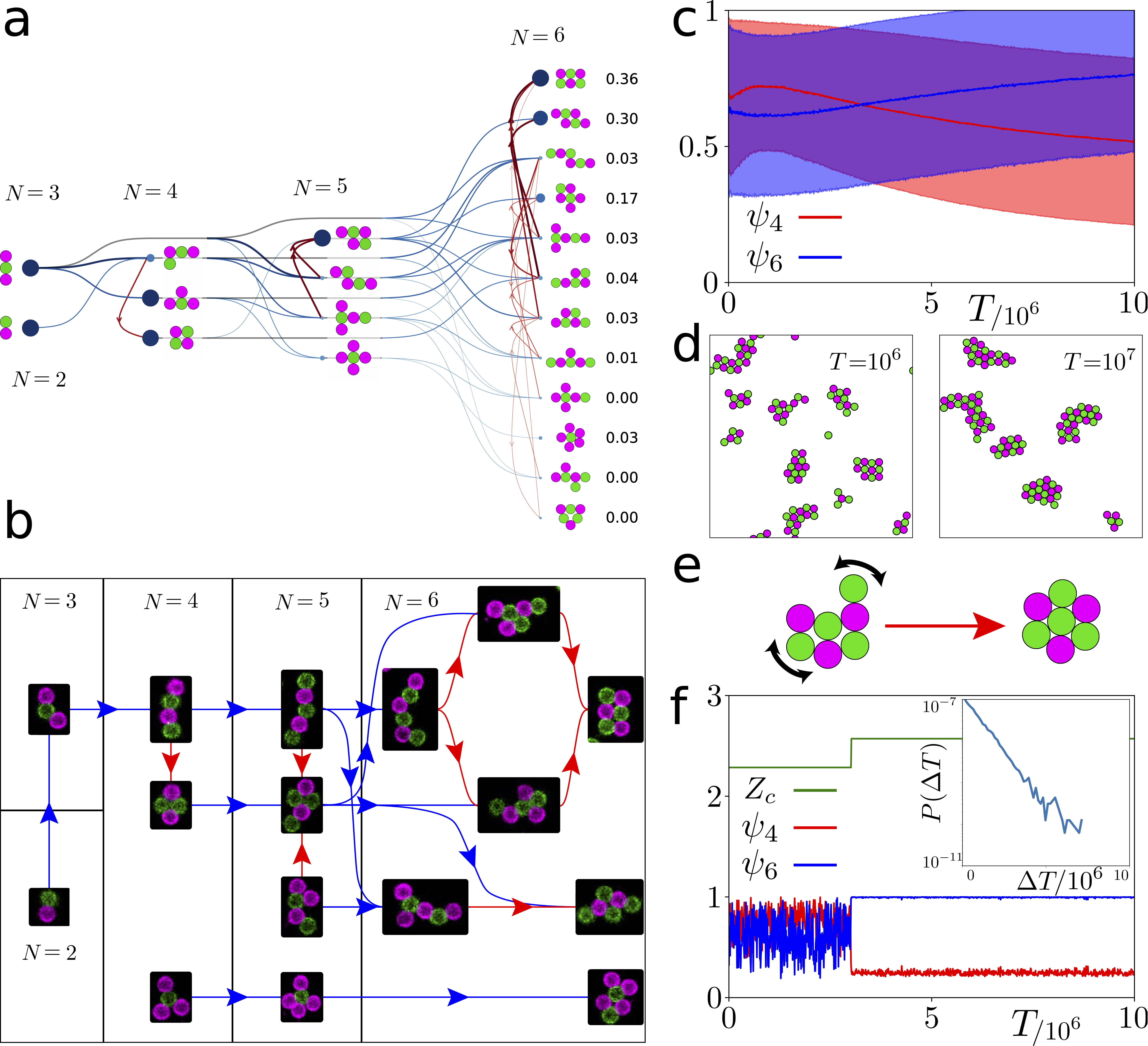}
		\caption{\small \textbf{Assembly pathways of flexible colloidal structures} (a) All possible  configurations and pathways for self-assembled structures obtained from spheres of equal size and number ratio up to $N=6$. Size and color of circles indicates relative likelihood of each aggregate. Probabilities of occurrence of a specific configurations are reported to the right of the structures for $N=6$. (b) Subset of most likely pathways resulting from addition of beads (blue arrows) or reconfiguration (red arrows) with exemplary confocal microscopy images from experiments. (c) Long time behaviour of the average four and six fold order parameter for systems of $N=100$ particles. (d) Snapshots of simulated clusters at short and long times. (e) Illustration of the transition between square and hexagonal arrangement of particles. (f) Simulated example of the transition outlined in (e). (f) (inset) The time taken for the transition shown in (d) is exponentially distributed. 
		}
		\label{fig:pathways}
	\end{figure}

	To illuminate the mechanism that can lead to square lattices, we consider the first bonds created on the path to making colloidal clusters. To do so, we construct all possible colloidal aggregates of a certain size ($N$) in which no beads can be removed from the structure without breaking a bond. The number of possible structures starts small but quickly diverges with increasing $N$. Fig.~\ref{fig:pathways}a shows examples of all possible structures containing $N\leq6$ for equal sized beads with equal number density. Note that the system is symmetric with respect to switching of the two populations of beads, and that we combined the two structures with opposite bead colors into one for clarity. By repeatedly simulating the assembly process, we can measure the relative likelihood of observing each structure, which is displayed as the size and color of the circle at each node of the diagram. Interestingly, there is only one possible aggregate of size $N\leq 6$ that is incompatible with a square lattice: the ``rosette'' where a central A particle is bound to five B particles. However, this configuration occurs only in 3\% of the cases for two reasons: first the space around the central A particle is almost full in this configuration meaning that the absorption of the fifth B particle has a low probability to occur\cite{chakraborty2022self}, second it requires a large imbalance of A and B type particles locally, which occurs with low probability for the equal number ratio's present in the experiment. Moreover, we find that the square unit we hypothesized to appear in chains of four or more particles to be present in 90\% of all clusters with size $N=6$. 
	
	If we assume that all bonds in a structure are formed irreversibly, we can simply draw every possible pathway of generating any specific aggregate; these pathways are shown as lines in Fig.~\ref{fig:pathways}a. The color of the lines indicates whether the bond represents the joining of two smaller clusters (blue) or an internal reconfiguration (red).
	From this information, we can construct the sub-network of pathways that accounts for over 95\% of aggregates and bonding events, see Fig.~\ref{fig:pathways}b. All of these configurations have been observed in experiments, which are shown as confocal microscope images in Fig.~\ref{fig:pathways}b. 
	
	The relative rate of adding a new particle (blue lines) and internal reconfiguration (red lines) depends on the age of the system. When the system is young there are many small clusters and the distance between them is low, thus bonds between clusters are relatively common. When the system is old, it is comprised of larger aggregates, which diffuse slower and the distance between them is increased. Therefore the timescale associated with the addition of beads increases over time and the structures have more time to reconfigure. For example, the most likely structures at $N=5$ and $N=6$ can only be formed by reconfiguration of less common structures, indicating that reconfiguration is fast compared to the addition of particles from an early stage. It is worthwhile to note that structures that are stable with respect to internal reconfiguration are the most common, because they can only be escaped from by addition of another particle or cluster which is governed by a longer time scale.  
	
	Our results suggest that the nature of the early assembly process leads to square bond networks with high probability, large lattices appearing at a later time will have at least small regions with a square network. This is reflected by an initial increase in the global average value of $\langle \psi_4\rangle$ for simulations of $N=100$ particles, see Fig.~\ref{fig:pathways}c. Square lattices have increased flexibility as there is space between the particles in the aggregate and the average connectivity is below the mechanical stability threshold\cite{mao2018maxwell}.
	
	At long times, however, we observe a decrease in $\psi_4$ and an increase in six fold order, as measured by the hexagonal order parameter $\psi_6=\langle\tfrac{1}{n_i}|\sum_{j}^{n_i} \exp (6i\theta_{ij})|\rangle_i$, indicating a crossover to hexagonal packing of the particles, see Fig.~\ref{fig:pathways}c (see SI Figure S5 for simulation images). This effect can be explained by the re-configurable nature of the bonds between the particles. A key feature of these floppy colloidal networks is the ability of the angles between the bonds to freely change, thus the square lattice has no resistance to shear deformations. These shear deformations result in a hexagonal packing of the particles and can become fixed with the creation of an additional bond. This is first possible for $N=7$ with the configuration shown in Fig.~\ref{fig:pathways}d. Indeed, upon reconfiguration and the formation of an additional bond, there is a predicted increase in both the average coordination number, $\langle Z_c\rangle:2.3\rightarrow2.6$, and the six fold order $\psi_6:0.65\rightarrow1.0$. At the same time, there is a predicted decrease in four fold order, $\psi_4:0.79\rightarrow0.21$ and the flexibility of the cluster. These predictions are closely in line with simulated results shown Fig.~\ref{fig:pathways}e. The time taken for this process is exponentially distributed with a well defined rate which indicates it arises from a Poisson process, see Fig.~\ref{fig:pathways}e (inset). Critically, this transition also corresponds to the cluster becoming completely rigid and all flexibility is lost as the system is left in a close packed state. 
	Thus, we conclude that while rearrangements are at the heart of the square lattice symmetry initially, they also can lead to a loss of it by bond formation at long times.

	\subsection{Semi-flexible networks with cubic beads}
\begin{figure}[htp]
		\centering
		\includegraphics [width=0.99\textwidth]{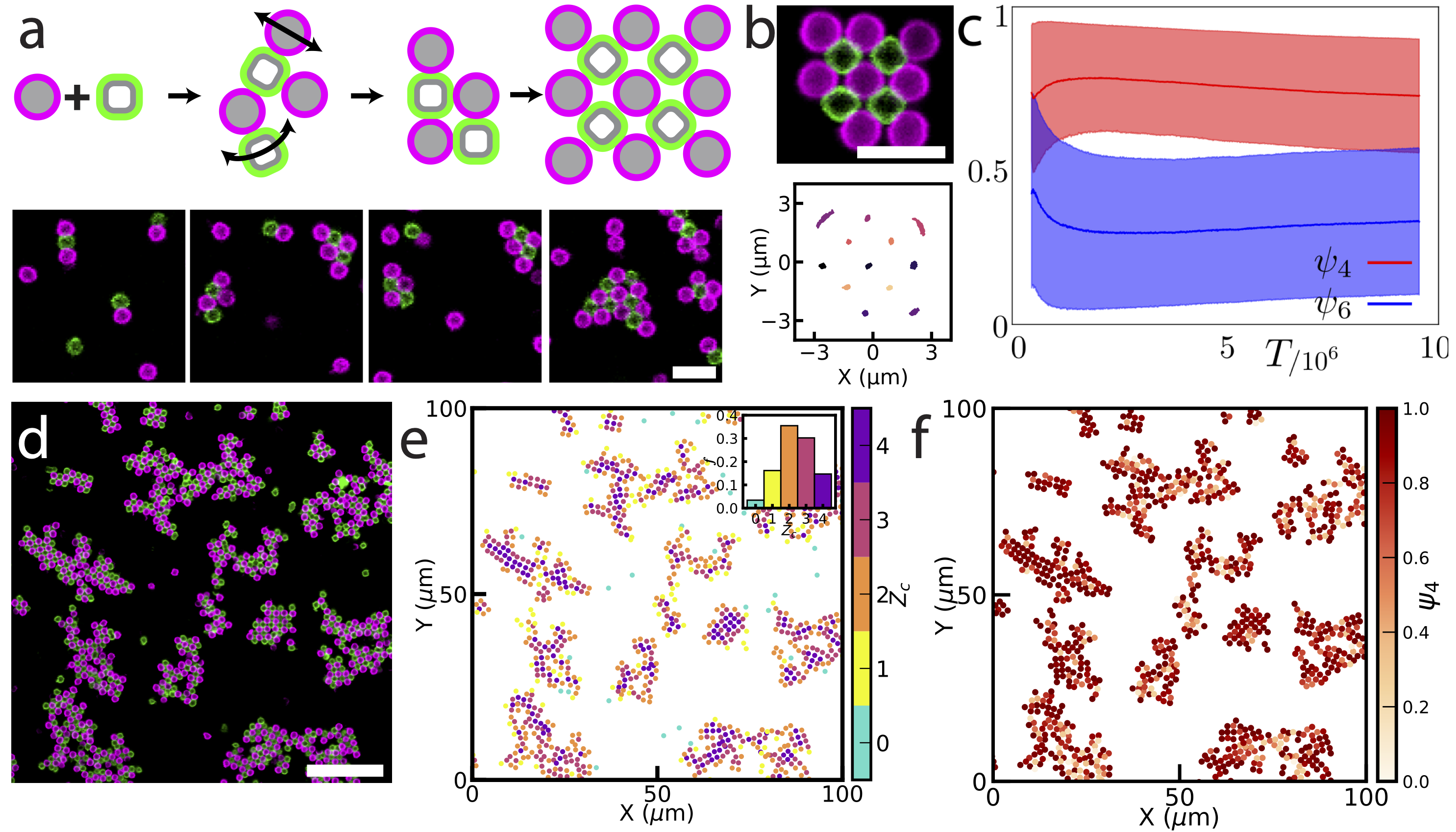}
		\caption{\small \textbf{Self-assembling flexible colloidal lattices from cubic and spherical particles} (a) (top row) Schematic and (bottom row) experimental confocal images showing the early stages of assembly of flexible lattices consisting of cubic (green) and spherical (magenta) particles. While spheres are confined to only move on the faces of the cubes, cubes can explore the full surface of a sphere. Scalebar is 5 $\mu$m.  (b) An exemplary confocal microscope image of a self-assembled square lattice and trajectories of individual particles within the lattices for 100 sec from which rotational and translational motions have been subtracted demonstrating the reduced flexibility when cubic beads are introduced. (c) Simulated long time behaviour of the four and six fold order parameter for systems of $N=100$ particles in which one population of beads is cubic. (d) Confocal microscopy image of self-assembled lattices of spheres and cubes obtained after 16 hours, see also SI Figure S2. Scalebar is 20 $\mu$m. The same snapshot is shown colored by (e) coordination number, $Z_c$, and (f) fourth order parameter, $\psi_4$. The inset of (e) shows the distribution of the coordination number.}
		\label{fig:cubes}
	\end{figure}

	To enhance the long-term stability by preserving the square symmetry we turn to cubic particles in the experiment, see Fig.~\ref{fig:cubes}. Because of their anisotropic shape, spheres have a strong preference to locate on the  flat surfaces after binding as more DNA linkers can bond there \cite{chakraborty2017colloidal, shelke2023flexible}. This thus imposes a four-fold symmetry for structures formed in 2D, which has been shown to allow selective assembly of clusters with four bounded spheres at highly asymmetric number ratios.\cite{shelke2023flexible} 
	
	Upon mixing cubic particles (1.35 $\mu$m side-length) and spheres (1.66 $\mu$m diameter) at approximately equal number ratios, we once again see the development of small scale structures with square symmetry in the early stages of assembly, (see \ref{fig:cubes}a and SI Movie 4 and 5) for consecutive assembly steps (see \ref{fig:cubes}d) for the results after $\approx$16 hours. Clearly, the presence of the cubic particles enhances the appearance of square order in the system.  
	
	The cubic particle shape not only guides the spheres into a square arrangement but also limits the reconfigurability of the formed structure. Once a sphere is bound to a cube, the sphere is unable to transition to another face of the cube. This reinforces the square symmetry in the assembling structures, see Fig.~\ref{fig:cubes}a. While the cube can freely rotate around the sphere as it has no edges, the flexibility of larger structures is still reduced relative to structures assembled from spheres only, see Fig.~\ref{fig:cubes}a, b. 
	
	The effect of the cubic shape is also reflected in the quantitative measurements: we find the coordination number to be smaller and $\psi_4$ to be higher than for aggregates with equal sized spheres, see Fig.~\ref{fig:cubes}d-f. The coordination number peaking below $Z_c=4$ stems in part from the fact that coordination by four spheres is the maximum possible value for a cubic particle. In addition, the reduced reconfigurability of the clusters containing cubic particles leads to them having more extended shapes and voids, resulting in more edges and corners and an estimated boundary ratio closer to $\gamma \approx 0.75$. Yet, the square order parameter is higher, showing the strong guiding effect of the square particle shape. Indeed, simulations show there is no crossover to $\psi_6$, see Fig.~\ref{fig:cubes}c and SI Figure S5, again confirming that the cubic shape strongly guides the assembly into a square symmetry and prevents reconfiguration towards hexagonal symmetries.  
	
	Thus, by introducing cubic particles, it is possible to influence the system to adopt square lattices. However, this is achieved at the cost of restricting the flexibility of the lattices. 
	
	\clearpage

	\subsection{Effect of breaking symmetry between two particle types}
\begin{figure}[t!]
		\centering
		\includegraphics [width=0.99\textwidth]{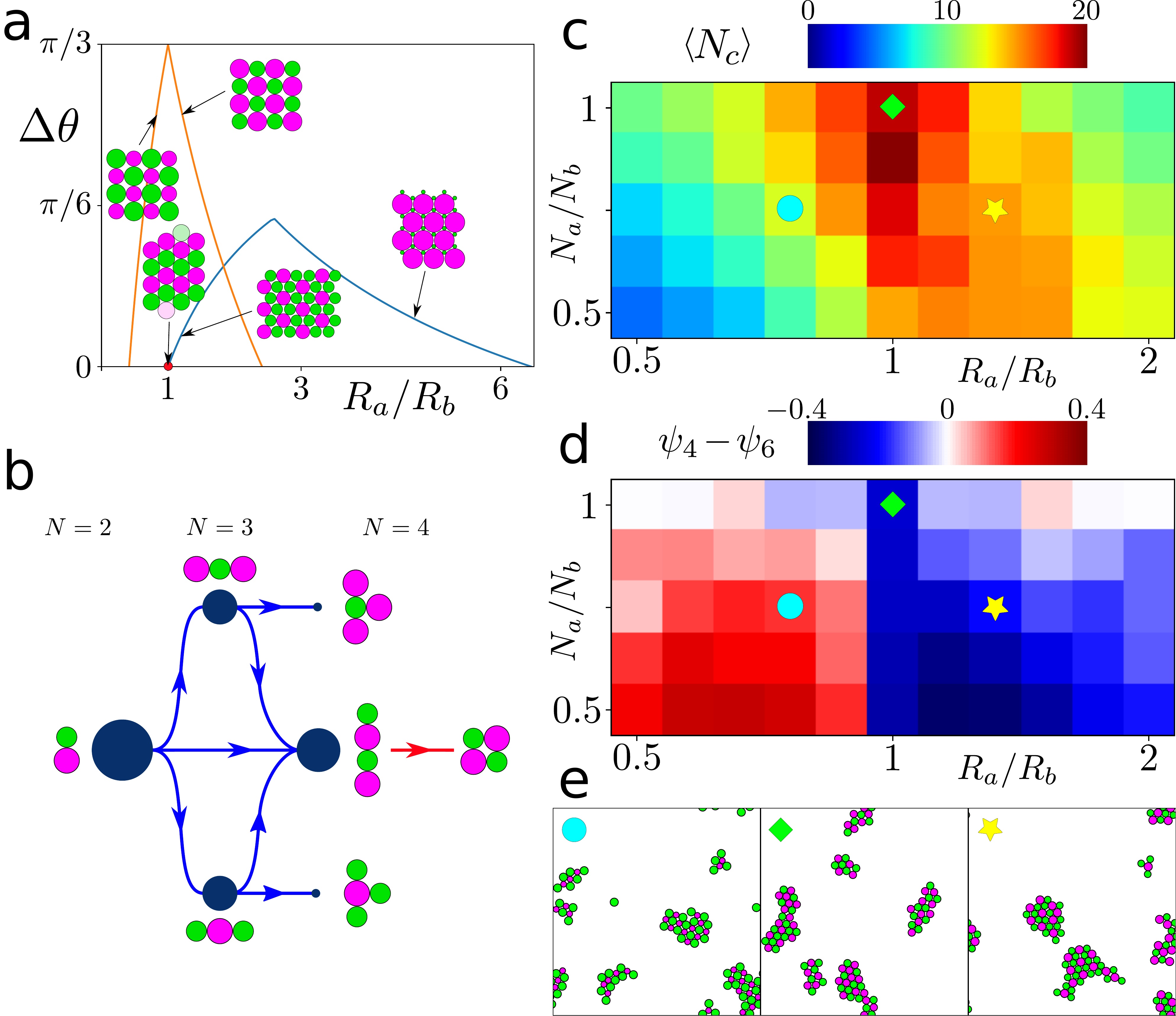}
		\caption{\small \textbf{Breaking symmetry between the two spherical particles of different size} (a) Networks with a square (orange) and hexagonal (blue) symmetry of the constituent particles and their maximum angular range as a function of particle size ratio. (b) Assembly pathways for clusters up to $N=4$. (c) Average cluster size and (d) $\psi_4-\psi_6$ for a simulation of $N=100$ particles in the long time limit, and (e) snapshots of simulations for each region of the parameter space. Parameter values are indicated by the symbols.  
		}
		\label{fig:ratios}
	\end{figure}

	The range of flexibility of a cluster depends on the bond network arrangement as well as the available space between the particles. The angular range of motion available for a perfect square lattice can be calculated using geometric arguments and depends on the aspect ratio $R_a/R_b$ of the particles with radii $R_a$ and $R_b$, see Fig.~\ref{fig:ratios}a (orange line). The square lattice is most flexible when the beads have equal size, but is susceptible to the transition towards a hexagonally symmetric structure, as outlined in Fig.~\ref{fig:pathways}d and shown as a red dot in Fig.~\ref{fig:ratios}a. Since hexagonal packing fills space optimally, there is no available space and flexibility is completely lost. However, when different particle sizes are used, the hexagonal lattice obtains a small range of motion. In contrast to the square lattice structures, such a hexagonal lattice is possible for a large range of size ratios, but has a smaller range of motion and requires a two to one ratio of beads to realize the AB$_2$ arrangement with $Z_c^{\rm{lat}} = 5$, see Fig.~\ref{fig:ratios}a (blue line). Therefore the flexibility and likelihood of a certain structure depends on the relative size and abundance of the constituent particles. 
	
	To understand how these two parameters can be tuned to target a certain lattice structure with a desired symmetry and flexibility, we again consider the assembly pathways that lead to different structures. The results presented in Fig.~\ref{fig:pathways}a correspond to systems that are symmetric with respect to exchange of the two different types of particle. If this symmetry is broken, for example by a size difference, the number of distinct combinations, and possible pathways between them, greatly increases. Fig.~\ref{fig:ratios}b shows the aggregation pathways for up to $N=4$ when the A and B particles are not identical. 
	
	The probability of taking any particular path is given by the product of the probability that an incident particle is of the correct type and that the particle can bind. 
	The probability that the incident particle has the correct type is simply related to the relative abundance of the two particle types. For example, the top configurations in Fig.~\ref{fig:ratios}b contain more magenta particles, thus are more likely when magenta particles are more abundant than green ones. The probability an incident particle is able to bind depends on the current configuration and aspect ratio and can be estimated from geometric arguments~\cite{chakraborty2022self}. In general, incident small particles are more likely to bind as the complementary large particles in a cluster present more surface area for them to bind to. For example, the top configurations in Fig.~\ref{fig:ratios}b all feature multiple large beads bound to a central smaller bead. Since there is less space on the surface of the smaller bead, this configuration is less likely than the inverse, in which multiple small beads bond to a central larger bead. 
	
	We measure the average cluster size, $\langle N_c\rangle$ after $10^7$ simulation steps as a proxy for cluster growth rate. We see that clusters grow fastest when the spheres are of equal size and abundance, see Fig.~\ref{fig:ratios}c. The growth rate is lowest when there is an abundance of larger particles ($R_a/R_b \approx N_a/N_b < 1$). In this scenario incident particles to a cluster are more likely to be larger, which are less likely to find a place to successfully bind. 
	
	We next assess how the symmetry of the lattice structure in the long time limit depends on the number and size ratio of the beads by measuring $\psi_4$-$\psi_6$, see Fig.~\ref{fig:ratios}d where positive values (red) indicate square structure and negative values (blue) indicate hexagonal structure. As we observed in Fig.~\ref{fig:pathways}c, for equal size and number ratio, in the long time limit the system adopts a predominantly 6 fold symmetric structure, see Fig.~\ref{fig:ratios}d,e (green diamond).
	When smaller beads are more abundant $\psi_6$ dominates and we observe hexagonal structures, see Fig.~\ref{fig:ratios}d,e (yellow star). When larger beads are more abundant $\psi_4$ dominates and we observe square structures, see Fig.~\ref{fig:ratios}d,e (blue circle). The regions with square structure show a corresponding drop in the coordination number (Fig.~\ref{fig:ratios}c, although this is also partly a function of the reduced cluster size which implies higher values of $\gamma$ (see SI).
	
	The mechanism by which $\psi_4$ is selected for can be understood by considering Fig.~\ref{fig:ratios}b. When the larger beads are more abundant the number ratio favors taking the uppermost path, however, entropic effects stemming from the size ratio penalize taking it. Thus, pathways in the middle are preferred, where square lattices are more common. However, since the spheres are not equal in size, the square lattices are not susceptible to the formation of additional bonds which ridigify the structure upon thermal-fluctuations, similar to  Fig.~\ref{fig:pathways}c,d,e. Thus by carefully controlling the size and number ratio of the beads, we can favor square lattices that retain a high degree of flexibility (see SI Figure S4).

\begin{figure}[b!]
		\centering
		\includegraphics [width=0.99\textwidth]{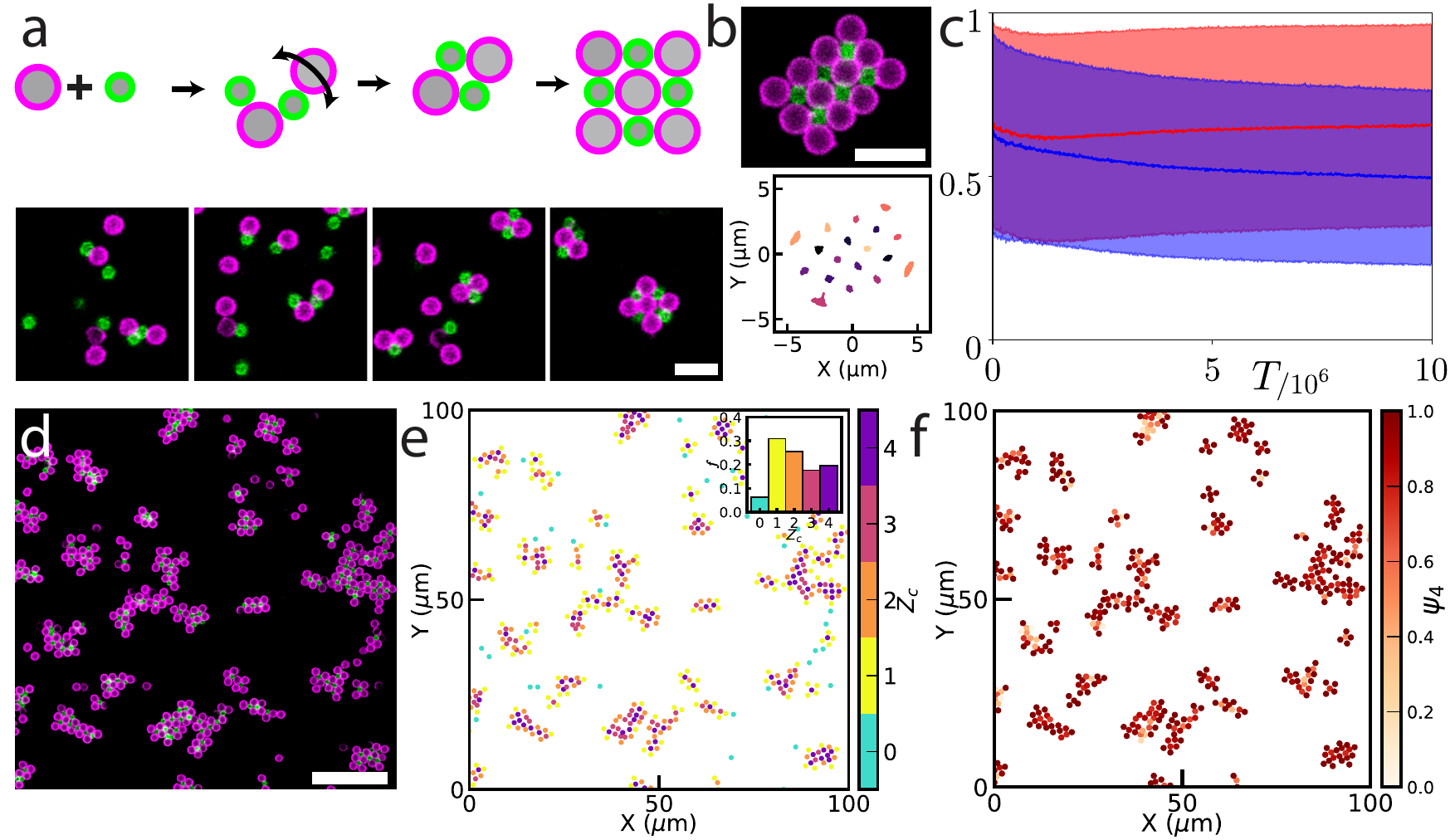}
		\caption{\small \textbf{Assembly pathways of binary mixtures of colloidal spheres with size ratio $R_a/R_b = 1.67$ and number ratio $N_a/N_b = 1.2$.} (a) Schematic (top) and confocal microscopy snapshots (bottom) of the assembly of flexible colloidal clusters consisting of 2.06$\pm$0.05 $\mu$m spheres (displayed in magenta) and 1.25$\pm$0.05 $\mu$m  spheres (displayed in green). Scalebar is 5 $\mu$m. (b) Confocal microscopy image of a self-assembled square lattice and trajectories of individual particles within the lattice obtained over 100 sec from which rotational and translational motions have been substracted to show the relative motion. Scalebar is 5 $\mu$m. c) Simulated long-time behavior of average $\Psi_4$ and $\Psi_6$ for a system of $N=100$ particles. (d) Confocal microscopy image of self-assembled lattices obtained after 16 hours. Scalebar is 20 $\mu$m. (e,f) The same snapshot  colored by (e) coordination number $Z_c$ and (f) square order parameter $\Psi_4$. The inset of (e) shows the distribution of the coordination number. 
		}
		\label{fig:unequal_size}
	\end{figure}

	We now apply our understanding to generate flexible and stable square lattices in the experimental system. In line with our results from Fig.~\ref{fig:ratios}, we select a size ratio $R_a/R_b = 1.67$ and number ratio $N_a/N_b = 1.2$ and overall area fraction of $\approx$$20\%$ to retain high bond flexibility and steer the assembly towards lattices with a high square symmetry. We employ complementary functionalized spheres with diameters $2.06\pm0.05$ $\mu$m and $1.25\pm0.05$ $\mu$m that can fully rearrange to form square lattices, see Fig.~\ref{fig:unequal_size}a. 
	
	Similar to previous experiments, we quickly observe the appearance of small structures with a square symmetry. These structures preserve flexibility as shown in Fig.~\ref{fig:unequal_size}b and SI Movie 9. After 15 min, we find that many small cluster have formed, (see SI Movie 7 and 8) Fig.~\ref{fig:unequal_size}d and SI Figure S3. The small structures are indicative of the reduced growth rate of clusters when the larger beads are more abundant and leads to a high estimated boundary ratio $\gamma\approx 0.8$, see SI Movie 8. We also observe a reduced coordination number, Fig.~\ref{fig:unequal_size}e, which is a reflection of both the reduced cluster size and the predominantly square lattice structures and increased value of $\psi_4$,see Fig.~\ref{fig:unequal_size}f. A simulation of the system with size ratio $R_a/R_b = 1.67$ and number ratio $N_a/N_b = 1.2$ shows that $\psi_4$ remains higher than $\psi_6$ even in the long time limit, Fig.~\ref{fig:unequal_size}c and SI for details.

	\clearpage
	\section{DISCUSSION}
	In this study, we demonstrated how guided self assembly can be utilized to generate flexible colloidal lattices with specific symmetries. Our findings indicate that when lattices are formed from spherical beads of equal-sized, they typically form flexible clusters with square symmetry. However, they can exhibit a spontaneous, irreversible transition to a lattice with hexagonal symmetry which results in a complete loss of flexibility. 
	One way that this transition can be avoided is by introducing cubic particles, which guides the assembly towards a four-fold symmetry but at the cost of flexibility. 
	%By calculating the local bond order parameter, we determined that the combination of geometric constraints and confined flexibility in the system of spheres and cubes yields the most locally ordered lattice structures, however flexibility is limited resulting from the confined movement of spheres over the cubes' surfaces. 
	We thus used a binary mixture of spheres instead to demonstrate that by controlling the relative size and abundance of the functionalized colloidal particles it is possible to influence the aggregation process of the lattices. We used these insights to assemble lattices with square symmetry from spherical beads, which can be tuned to retain a high degree of flexibility and be stable in the long time limit. 
	
	Our findings demonstrate that particle shape, size and number ratio, as well as reconfigurability all play crucial roles in the pathway towards and the time required for the formation of flexible lattices. These results on identifying an optimal pathway towards flexible colloidal structures may serve as a valuable tool for selecting the appropriate building block shape, size and motion range to design flexible colloidal materials with customized properties. 
	The resulting lattices are exciting models for testing predictions about floppy modes in thermal systems\cite{broedersz_modeling_2014} and may be used as a starting point to create functional colloidal structures that, for example, execute work by changing their conformation.\cite{brandenbourger_limit_2022} 
	
	\section{MATERIALS AND METHODS}
	
	\subsubsection{Materials}
	Silica microspheres of diameters  1.25$\pm$0.05 $\mu$m, 1.66$\pm$0.05 $\mu$m, 2.06$\pm$0.05 $\mu$m,  in 5 wt/v\% suspension were purchased from Microparticles GmbH. Sodium chloride, Ethanol, Sodium hydroxide, Ammoniun hydroxide (28-30 v/v\%),  Iron(III) Chloride Hexahydrate (FeCl$_3 \cdot $6H$_2$O), Tetraethyl orthosilicate (TEOS), 4-(2-hydroxyethyl)-1-piperazineethanesulfonic acid (HEPES), trimethoxysilyl propyl methacrylate (TPM), were purchased from Sigma-Aldrich and used as received. 1,2-dioleoyl-\textit{sn}-glycero-3- phosphocholine (DOPC), 1,2-dioleoyl-\textit{sn}-glycero-3-phosphoethanolamine-N-\-[methoxy(po-lyethylene glycol)-2000]  (DOPE-PEG2000), 1,2-dioleoyl-\textit{sn}-glycero-3-phospho\-ethanol\-amine-N-(lis-
	samine rhodamine B sulfonyl)  (DOPE-Rhodamine) and dye 23(dipyrrometheneboron difluoride)-24-nor-cholesterol (TopFluor-Cholesterol) were obtained from Avanti Polar Lipids, Inc.. We used Milli-Q water for all experiments. DNA strands were purchased from Eurogentec. The sequences of DNA used were: \\
	Strand A: Double Stearyl-HEG-5$^\prime$-TT-TAT-CGC-TAC-CCT--TCG-CAC-AGT-CAC-CTT-CGC-ACA-GTC-ACA-TTC-AGA-GAG-CCC-TGT-CTA-GAG-AGC-CCT--GCC-TTA-CGA-\textit{G-T-A-G-A-A-3$^\prime$-ATTO488}, \\
	Strand B: Double Stearyl-HEG-5$^\prime$-TT-TAT-CGC-TAC-CCT--TCG-CAC-AGT-CAC-CTT-CGC-ACA-GTC-ACA--TTC-AGA-GAG-CCC-TGT-CTA-GAG-AGC-CCT--GCC-TTA-CGA-\textit{T-T-C-T-A-C-3$^\prime$-Cy3}, \\
	Strand C: 5$^\prime$-TCG-TAA-GGC-AGG-GCT-CTC-TAG-ACA-GGG--CTC-TCT-GAA-TGT-GAC-TGT-GCG-AAG-GTG--ACT-GTG-CGA-AGG-GTA-GCG-ATT-TT-3$^\prime$, \\
	Strand D: Double Stearyl-5TT-TAT-CGC-TAC-CCT-TCG-CAC-AGT-CAA-TCT-AGA-GAG-CCC-TGC-CTT-ACG-A  and Strand E: TCG-TAA-GGC-AGG-GCT-CTC-TAG-ATT-GAC-TGT-GCG-AAG-GGT-AGC-GAT-TTT. The linker sequence in strand A and strand B is italicised. 
	
	\subsubsection{Synthesis of silica cubes.}
	Hematite cubes of edge length 1.08±0.04 $\mu$m were  synthesized following reference\cite{sugimoto1992preparation} and coated with a 0.135 $\mu$m silica layer using the process described in \cite{wang2013shape} which resulted in hematite-silica core-shell cubes of edge length  1.35±0.07 $\mu$m. In a typical synthesis of hematite cubes, 100 ml of aqueous 2M FeCl$_3 \cdot $6H$_2$O were prepared in a 500 ml Pyrex bottle. Next, 100 ml of 5M NaOH solution were added while stirring at 1000 rpm for 50 seconds. Then, the mixture was stirred continuously for another 10 minutes and subsequently placed in a preheated oven and left undisturbed at 100 $^\circ$C for 8 days. The resulting hematite cubes were washed several times using centrifugation and redispersion in milliQ water. To coat them with a thick 0.135 $\mu$m layer of silica, 100 ml of ethanol and 0.6 g of synthesized cubes were mixed under ultrasonication and mechanical stirring in a 2-neck round bottom flask at 50 $^\circ$C. Subsequently, 5 ml of water, 15 ml of ammonium hydroxide solution and 0.6 ml of TEOS were poured into the reaction flask and allowed to coat the particles' surface with silica layer for 5hr under the same conditions. The resulting silica coated cubic particles were first washed three times with ethanol and then three times with water to remove unreacted chemicals by repeated centrifugation and redispersion. To remove the hematite core, 5ml of silica coated hematite cubes was dispersed in 5 ml 37 wt/v\% HCl solution. The HCl was allowed to etch the hematite cubes for 12-15 hours. After complete etching as confirmed by bright field microscopy, the resulting hollow silica cubes washed with water by centrifugation and redispersion three times.
	
	\subsubsection{Preparation of small unilamellar vesicles (SUVs)}
	Small unilamellar vesicles (SUVs) were prepared using a  protocol described in reference\cite{rinaldin2019colloid}. For the preparation of SUVs, we used 77 $\mu$l of 25 g/L DOPC, 7.34 $\mu$l of 10 g/L DOPE PEG 2000, and 2 $\mu$l of either 1 g/L dye DOPE-rhodamine or 2 $\mu$l of 1 g/L TopFluor-Cholesterol dissolved in chloroform were mixed together in a glass vial. Subsequently, the lipid mixture was dried for at least 2 hrs in a desiccator (Kartell) attached to a vacuum pump (KNF LABOPORT N816.3KT.18). Then, 1 ml of buffer solution consisting of  50 mM NaCl and 10 mM HEPES at pH 7.4 was added to the dried lipid. The prepared solution was vortexed for 30 min during which the solution became turbid indicating the formation of giant multilamellar vesicles. The dispersion of giant multilamellar vesicles was extruded with a Avanti Polar Lipids mini extruder 21 times through a 50 nm polycarbonate membrane supported with filter paper to achieve SUV formation. The prepared SUVs were stored in the fridge at 4 $^\circ$C and used for up to 3 days.  
	
	\subsubsection{DNA Hybridization}
	We used double stranded DNA  with a complementary 6 base-pair single stranded end to which we refer to as "linker" and inert double-stranded DNA strands for bonding and stabilizing the colloids, respectively. Prior to use, single strand DNA was hybridized with the complementary backbone to obtain linker DNA. Strand A was hybridized with strand C to yield double-stranded linker DNA, strand B with strand C to obtain the complementary double-stranded linker DNA, and strand D with strand E to create double-stranded inert DNA (DNA strands are listed in the materials section). For hybridization, we typically mixed 10 $\mu$l of 20 $\mu$M single strands and 10 $\mu$l of 20 $\mu$M complementary backbone in 90 $\mu$l buffer (200mM NaCl, 10mM HEPES, at pH 7.4) solution. The DNA solutions were placed in a preheated oven at 94 $^\circ$C for 30 minutes. The oven then was switched off and allowed to cool slowly overnight. After cooling, the hybridized DNA strands were stored at 4 $^\circ$C and used for up to 2 months.

	\subsubsection{Functionalization of Colloidal Particles with a Lipid Bilayer Containing Linker and Inert DNA}
	To coat particles with a lipid bilayer we used a 25:1 surface ratio of SUVs to particles. We maintained the same surface area ratio of SUVs to particles when coating differently sized particles. Typically, for 2 $\mu$m particles, we use 100 $\mu$l of 0.25 wt/v\% particles in Milli-Q water and  mixed them with 37 $\mu$l SUVs. Then, the dispersion was rotated at 8 rpm for 1h. During this period, SUVs collide, burst, spread and form a bilayer on the particles' surface.  Then, the coated particles were centrifuged at 800 rpm for 2-5 min and the supernatant containing excess SUVs was removed using a micropipette. Subsequently, linker DNA was added to the particles at a concentration of $2\times10^{4}$ labeled with dye DOPE-rhodamine (magenta) and $4\times10^{4}$ labeled with dye  TopFluor-Cholesterol (green), respectively. We employed three combinations of particles: 1) equal sized spheres of 2.06$\pm$0.05 $\mu$m diameter, 2) unequal sized spheres of 2.06$\pm$0.05 $\mu$m and 1.25$\pm$0.05 $\mu$m diameter, and 3) 1.66$\pm$0.05 $\mu$m spheres and 1.35 $\pm$0.07 $\mu$m cubes. Inert DNA at a nominal concentration of $5\times10^{4}$ DNA strands/$\mu$m$^2$ was added to all particles suspensions. The suspensions were then rotated for another 1h. Thereafter, each suspension was centrifuged and washed 2 times with a 50 mM NaCl HEPES buffer. 
	
	\subsubsection{Sample Preparation and Imaging} 
	For all self-assembly experiments, 2.06$\pm$0.05 $\mu$m  spheres and 2.06$\pm$0.05 $\mu$m  spheres a number ratio of 1:1, 2.06$\pm$0.05 $\mu$m  spheres and 1.25$\pm$0.05 $\mu$m  spheres a number ratio of 1.2:1, 1.66$\pm$0.05 $\mu$m  spheres and 1.35 $\pm$0.07 $\mu$m  cubes a number ratio of 1:1 was maintained. Polyacrylamide (PAA) coated coverslips were used as substrates and coated with PAA by following the protocol in ref \cite{verweij2020flexibility}. In a 0.5 ml vial, the amount of spheres that was calculated to yield an approximately 10\% area fraction after moving to the sample chamber of 2.06 ± 0.05 $\mu$m DNA functionalized and 10\% area fraction of complementary DNA functionalized spherical particles were mixed. Additionally, 200 µl of a 50 mM NaCl and 10 mM HEPES buffer, pH 7.4 was added to particles. The mixture was then transferred to a customized sample holder and allowed to self-assemble and while being captured with brightfield or confocal microscopy over time upto 2 hrs. During self-assembly the sample temperature was maintained at 23.0$\pm$0.7 $^\circ$C using a temperature controlled water bath (Julabo) equipped with a Pt100 sensor that was inserted into the sample cell. Images and videos were captured with a  Nikon inverted TI-E microscope equipped with an A1 confocal scan head and Prime BSI Express camera (Teledyne Photometrics) for capturing the sample evolution in brightfield mode. The images were taken with a 60x oil objective (N.A. 1.4) at frame rates of up to 25 fps.
	
	\subsubsection{Particles Detection and Image Analysis} 
	To extract the $x$ and $y$ coordinates from TIF images, the images were first processed by separating the green and magenta channels. A Hough Circle Detection technique was employed to identify particles appearing as bright circular rings within these channels. Following the detection of particle positions, the coordinates were stored in CSV files. These coordinates were subsequently used for analysis using custom Python scripts.
	
	\subsection{Simulation details}
	
	The simulation consists of $N$ beads in a square region with sides of length $L$ and periodic boundary conditions. Each bead belongs to either population $A$ or $B$ which have complimentary DNA tethers; there are $N_A$ ($N_B$) beads in population $A$ ($B$) such that $N= N_A+N_B$. The relative radii of the populations may also be distinct, with the radius of bead $i$ being described by $R_i$ with $R_i=R_A$ ($R_i=R_B$) for beads belonging to population $A$ ($B$). The packing fraction of beads in the simulation is therefore given by $\phi = (N_AR_A^2 + N_BR_B^2)\pi/L^2$.
	
	The location of bead $i$ is described by the vector $\underline{r}_i$, the motion of which is described by the following overdamped Langevin equation
	
	\begin{equation}
		\dot{\underline{r}}_i = \mu\sum_j \underline{F}_{ij} + \eta\underline{\hat{\xi}}
		\label{eq:langevin}
	\end{equation}
	
	\noindent where $\underline{\hat{\xi}}$ is an uncorrelated randomly orientated unit vector capturing thermal fluctuations, $\eta$ is the magnitude of the random movements due to thermal noise, $\mu$ is a mobility coefficient and $\underline{F}_{ij}$ captures the forces exerted on particle $i$ due to interactions with particle $j$. In principle, the coefficients could depend on which population particle $i$ belongs to but here we keep them constant for all particles. Equation~\ref{eq:langevin} is integrated according to $
	\underline{r}_i^{t+1} = \Delta \underline{\dot{r}}_i^t
	$
	to simulate the motion of the particles where $\Delta$ is the length of the time step. 
	
	The force between two particles arises from two effects, collisions and bond formation, such that $\underline{F}_{ij} = \underline{F}_{ij}^c + \underline{F}_{ij}^b$.
	
	The collision force between two particles is given by
	\begin{equation}
		\underline{F}_{ij}^c=\begin{cases}
			K^c\underline{\hat{r}}_{ij}(r_{ij} - (R_i + R_j)), & \text{if $r_{ij} < R_i + R_j$}\\
			0, & \text{otherwise.}
		\end{cases}
	\end{equation}
	where we have introduced $\underline{r}_{ij}$ as the vector from particle $i$ to particle $j$ such that $\underline{r}_{ij} = -\underline{r}_{ji}$ and $r_{ij} = r_{ji}$ as the distance separating the two particles; $\hat{}$ denotes a unit vector; $R_i$ describes the radius of particle $i$; $K^c$ the elastic collision parameter.
	
	The bonding force between two particles is given by 
	\begin{equation}
		\underline{F}_{ij}^b=\begin{cases}
			K^bA_{ij}\underline{\hat{r}}_{ij}(r_{ij} - (R_i + R_j + L_{\text{bond}})), & \text{if $r_{ij} > R_i + R_j + L_{\text{bond}}$}\\
			0, & \text{otherwise}
		\end{cases}
	\end{equation}
	Where we have introduced $L_{\text{bond}}$ as the relaxed length of a pair of bound DNA tethers, $K^b$ is the tether elastic coefficient and $A_{ij}$ is the adjacency matrix. 
	
	The adjacency matrix describes which particles are bound and hence defined as
	\begin{equation}
		A_{ij} = \begin{cases}
			1 , & \text{if particle $i$ and $j$ are bound}\\
			0, & \text{otherwise}
		\end{cases}
	\end{equation}
	The adjacency matrix is updated dynamically as follows: If two particles with complimentary tethers come into contact ($r_{ij} < R_i + R_j + L_{\text{bond}}$), a bond is formed between them. If two bound particles become sufficiently separated ($r_{ij} > R_i + R_j + L_{\text{max}}$), the bond between them is broken.
	The adjacency matrix is symmetric and completely describes the graph of the bond network between the particles. Clusters in the simulation are connected sub-graphs of the whole system and can be represented by a sub matrix of $A_{ij}$. To identify a cluster, it is sufficient to generate the eigenvalues of its adjacency matrix. This can be used to track the existence of specific clusters in the simulation and generate the data shown in Fig.~\ref{fig:pathways}a.
	
	Without loss of generality, we redefine $\eta = \eta/\Delta$, $K^c = K^c\mu/\Delta$, $K^b = K^b\mu/\Delta$ and set their values at $\eta = 10^{-2}$ and $K^b = K^c = 0.45$. All simulations are performed with fixed packing fraction $\phi = 0.1$ which defines the length of the periodic boundaries; this helps de-correlate the system from its initial conditions as the particles have space to move before they bind. We set $L_{\text{bond}} = 0.04$ to mirror the scales observed in experiments. Finally we set $L_{\text{max}} = 10L_{\text{bond}}$ to close the equations. Unless otherwise stated, all simulations are initialized with random, non-overlapping particle positions and run for $T=10^7$ time steps.
	
	Results presented in Fig.~\ref{fig:pathways}a and c are based on $500$ simulations with $N_A=N_B=50$ particles with sizes $R_A=R_B=1$. 
	
	Cluster probabilities in Fig.~\ref{fig:pathways}a give the likelihood that a random cluster of given size has a specific configuration over the duration of the simulation. This is a time-independent quantity, thus it includes both the likelihood of a cluster forming and its stability. Short lived structures such as for example the N=6 ring possess a low probability. 
	
	Pathway probabilities in Fig.~\ref{fig:pathways}a give the likelihood of each pathway relative to all other pathways that result in a cluster of the same size. This is also integrated over the duration of the simulation. 
	
	Fig.~\ref{fig:pathways}e is based on a simulation of $N_A=3$, $N_B=4$, and $R_A=R_B=1$. The simulation is started with the configuration shown on the left of Fig.~\ref{fig:pathways}d. Fig.~\ref{fig:pathways}e (inset) is based on $1000$ realisations of this simulation.
	
	Fig.~\ref{fig:cubes}c is based on $500$ simulations with $N_A=N_B=50$ particles with sizes $R_A=R_B=1$. This simulation gives the same average coordination number as that in Fig.~\ref{fig:pathways}c and Fig.~\ref{fig:unequal_size}c for direct comparison. 
	
	Fig.~\ref{fig:ratios}c is based on simulations with $N = N_A+N_B = 100$ and $\min\{R_A,R_B\} = 1$. There are $25$ simulations at each point in the parameter space. 
	
	Fig.~\ref{fig:unequal_size}c is based on $500$ simulations with $N_A = 60$, $N_B = 50$, $R_A = 1.0$, $R_B = 0.6$. This simulation gives the same average coordination number as that in Fig.~\ref{fig:pathways}c and Fig.~\ref{fig:cubes}c for direct comparison. 
	
	\subsubsection{Simulating square particles}
	
	Square particles with side length $2R$ are approximated by four discs each with radius $R/2$ arranged in a non-overlapping square around the center of the particle. The center of each disc ($k$) representing square particle $i$ located at $\underline{r}_i$ is given by 
	\begin{equation}
		\underline{c}_k = \underline{r}_i + \sqrt{2}R\hat{\underline{\theta}}_i(k)/2
	\end{equation}
	where $\hat{\underline{\theta}}_i(k)$ is a unit vector given by the equation
	\begin{equation}
		\hat{\underline{\theta}}_i(k) = [\cos(\theta_i + k\pi/2),\sin(\theta_i + k\pi/2)]
	\end{equation}
	for $k\in\{0,1,2,3\}$. Here $\theta_i$ is the orientation of square particle $i$ relative to the $x$ axis.
	
	When calculating the collision forces with a square particle, we consider overlaps of the constituent discs. We introduce $\underline{F}_{ij}^c(k_i)$ to describe the force on the $k_i$th disc of the $i$th square due to collisions with particle $j$. Each collision can introduce a translational and a rotational force on the square particle $i$, thus we must treat the two components differently. 
	
	The translational component is given by:
	\begin{equation}
		\underline{F}_{ij}^c = \sum_{k_i}\hat{\underline{\theta}}_i(k).\underline{F}_{ij}^c(k_i)
	\end{equation}
	which can be simply introduced to equation~\ref{eq:langevin}. The resulting rotational component is given by:
	\begin{equation}
		\tau_{ij}^c = \sqrt{2}R\sum_{k_i}\hat{\underline{\theta}}_i(k)\times\underline{F}_{ij}^c(k_i)/2
	\end{equation}
	where we have taken the out of plane component of the cross product.
	
	This requires an additional overdamped Langevin equation to describe the evolution of the orientation of the square given by:
	\begin{equation}
		\dot{\theta}_i = \mu_r\sum_j\tau_{ij}^c + \eta_r
	\end{equation}
	Where $\eta_r$ is a random number sampled from a uniform distribution with range $[-\omega,\omega]$ and $\mu_r$ is a rotational mobility coefficient. For the simulations in Fig.~\ref{fig:cubes}c we set $\omega = 10^{-2}$ and $\mu_r = 2.5$.
	
	Between a square $i$ and a circular $j$ particle the collision force must be calculated for each disc in the square, with the force on each disc being:
	\begin{equation}
		\underline{F}_{ij}^c(k_i)=\begin{cases}
			K^c\underline{\hat{r}}_{ij}(k_i)(r_{ij}(k_i) - (R_i/2 + R_j)), & \text{if $r_{ij} < R_i/2 + R_j$}\\
			0, & \text{otherwise}
		\end{cases}
	\end{equation}
	Where $\underline{{r}}_{ij}(k_i)$ is now the vector from the center of particle $j$ to the center of disc $k_i$ around particle $i$. 
	
	When calculating the force between two square particles, the collisions between all eight discs need to be considered. This is a simple extension of the previous case given by:
	\begin{equation}
		\underline{F}_{ij}^c = \sum_{k_j}\sum_{k_i}\underline{F}_{ij}^c(k_i,k_j)
	\end{equation}
	With the force on each disc being:
	\begin{equation}
		\underline{F}_{ij}^c(k_i,k_j)=\begin{cases}
			K^c\underline{\hat{r}}_{ij}(k_i,k_j)(r_{ij}(k_i,k_j) - (R_i/2 + R_j/2)), & \text{if $r_{ij} < R_i/2 + R_j/2$}\\
			0, & \text{otherwise}
		\end{cases}
	\end{equation}
	Where $\underline{{r}}_{ij}(k_i,k_j)$ is now the vector from the center of disc $k_j$ around particle $j$ to the center of disc $k_i$ around particle $i$. 
	
	The bonding forces are not changed for the square particles. This recreates the greater bonding strength on the sides relative to the edges of the cubes.

	\section*{Supplementary materials}

	Supplementary Text\\
	Figs. S1 to S5\\

	\bibliography{main}
	
	\bibliographystyle{Science}

	\section*{Acknowledgments}
	Authors thank Rachel Doherty for help in SEM imaging of particles. We also thank Julio Melio for assistance in particles tracking. DJK gratefully acknowledges funding from the European Research Council (ERC Starting Grant number 758383, RECONFMAT). DJGP gratefully acknowledges funding from the Swiss National Science Foundation (SNSF Starting Grant TMSGI2\_211367).
	
	\subsection*{Funding}
	European Research Council (ERC Starting Grant number 758383, RECONFMAT)\\
	Swiss National Science Foundation (SNSF Starting Grant TMSGI2\_211367)\\
	
	\subsection*{Author contributions}
	Conceptualization: YS, DJGP, DJK 
	Methodology: YS, DJGP, DJK
	Software: DJGP
	Investigation: YS, DJGP
	Formal analysis: YS, DJGP, DJK
	Visualization: YS, DJGP
	Supervision: DJGP, DJK
	Writing—original draft: YS, DJGP, DJK
	Writing—review \& editing: DJGP, DJK
	Funding acquisition: DJGP, DJK
	
	\subsection*{Competing interest}
	Authors declare that they have no competing interests. 
	\subsection*{Data and materials availability}
	All data are available in the main text or the supplementary materials.

	\newpage
	\section{Supplementary Figures}
	\subsection{Spheres with equal size}
	\begin{figure}[h!]
		\centering
		\includegraphics [width=0.99\textwidth]{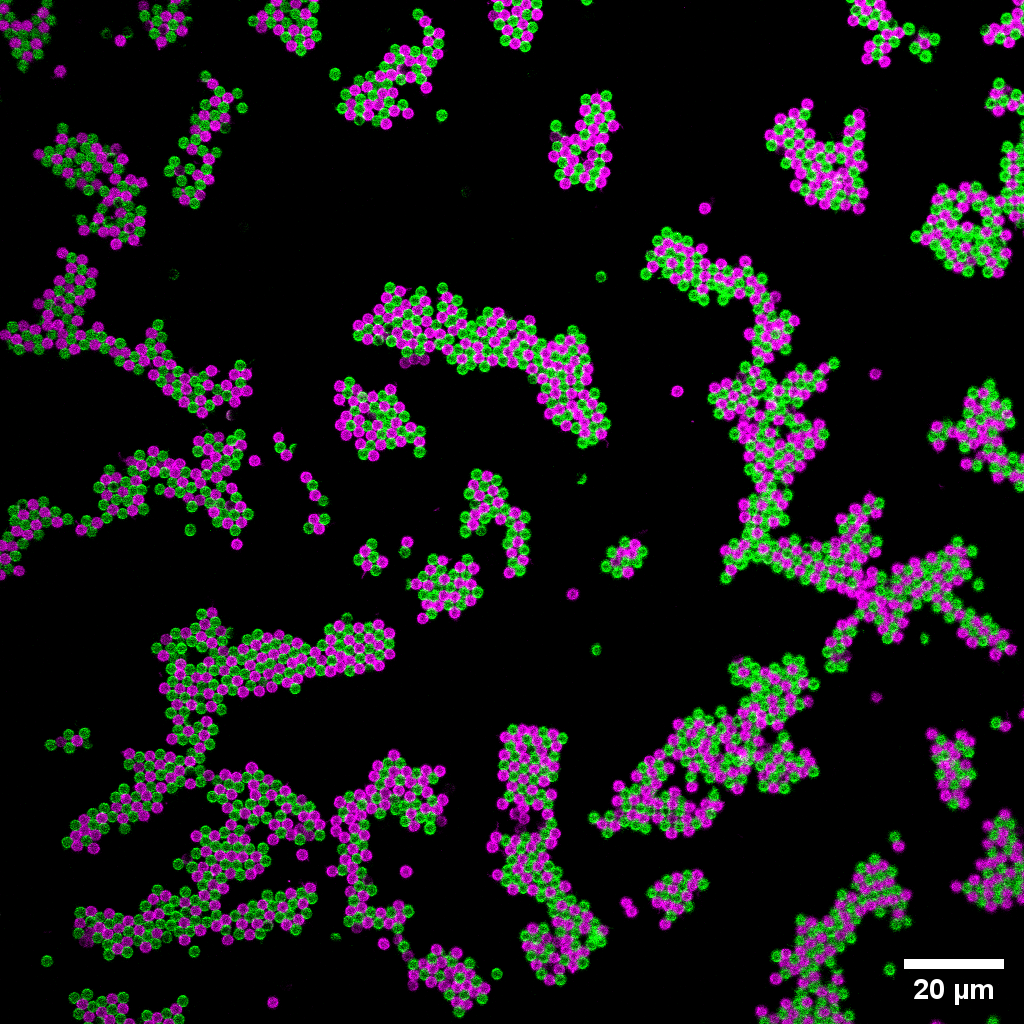} 
		\caption{Large field view confocal microscopy image of flexible colloidal lattices  assembled from 2.12$\pm$0.05 $\mu$m spheres functionalized with complementary DNA linkers (distinguishable by addition of different dyes displayed in magenta and green) as described in the main text and shown in Figure 1.}
		\label{Equal_size_sphere}
	\end{figure}
	
	\subsection{Self-assembly of spheres and cubes}
	
	\begin{figure}[h!]
		\centering
		\includegraphics [width=0.99\textwidth]{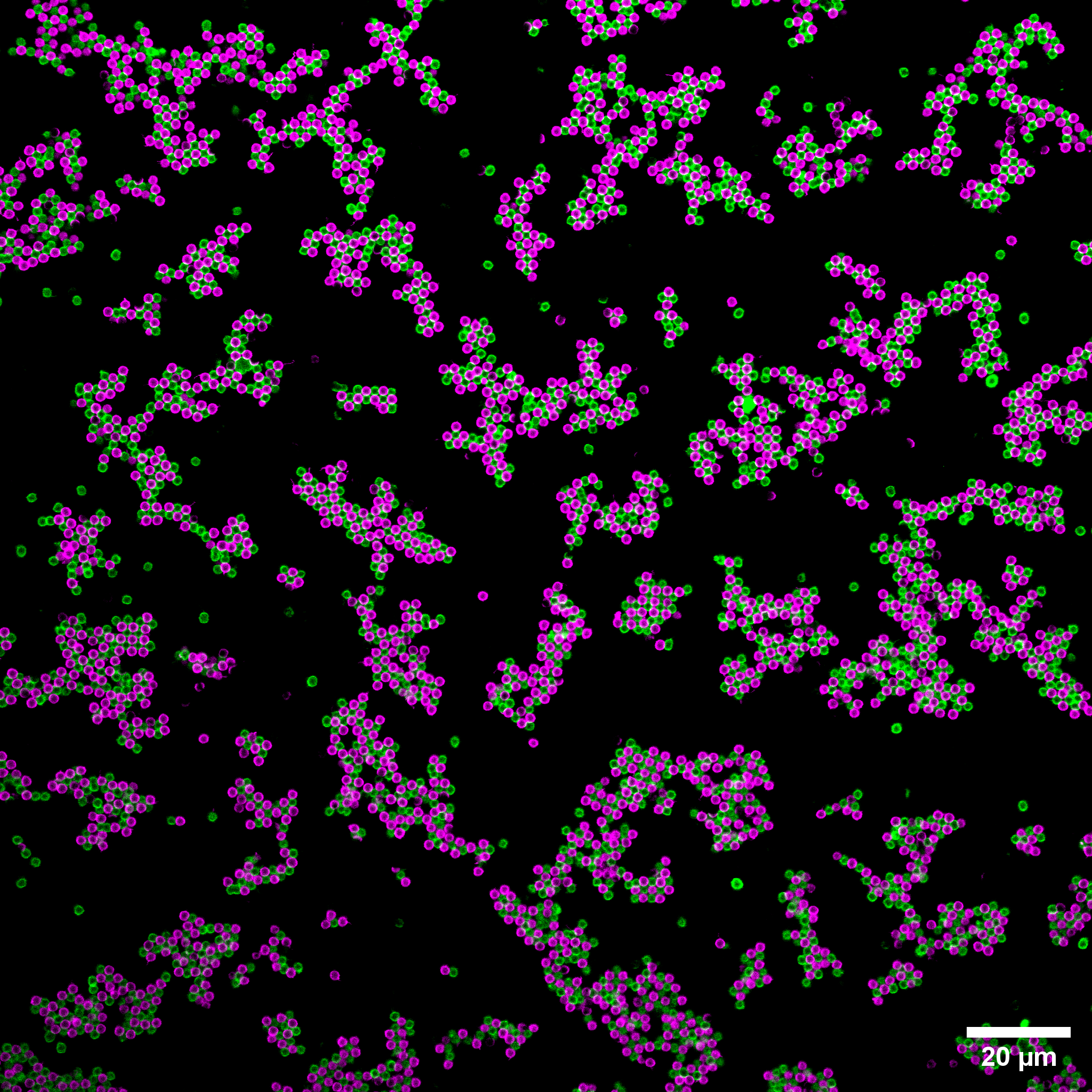} 
		\caption{Large field view confocal microscopy image of flexible colloidal lattices  self-assembled from 1.66$\pm$0.05 $\mu$m spheres (displayed in magenta) and 1.30$\pm$0.07 $\mu$m  cubes (displayed in green).}
		\label{Cube_large_view}
	\end{figure}
	\clearpage
	\subsection{Self-assembly of unequal sized spheres}
	
	\begin{figure}[h!]
		\centering
		\includegraphics [width=0.99\textwidth]{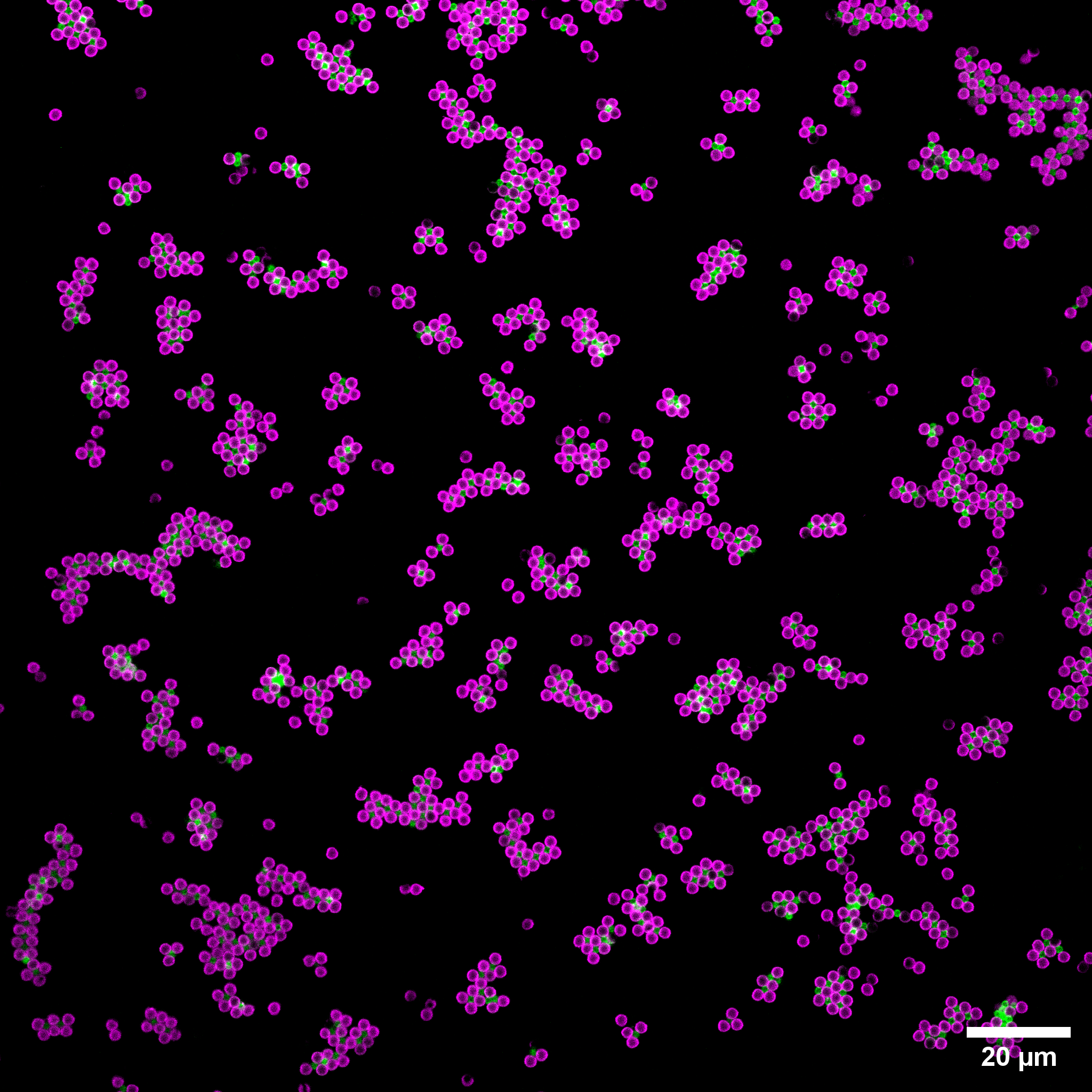} 
		\caption{Large field view confocal microscopy image of flexible colloidal lattices  consisting of 2.06$\pm$0.05 $\mu$m spheres (displayed in magenta) and 1.25$\pm$0.05 $\mu$m  spheres (displayed in green)}
		\label{Cube_large_view}
	\end{figure}
	
	\clearpage
	\newpage
	\section{Discussion on average coordination number}
	
	We calculate the coordination number of a particle, $Z_c$, as the number of bonds it has formed. The average coordination number of a system, $\langle Z_c\rangle$, is  then simply the average over all particles. For an infinite square lattice, all particles are bonded to their four nearest neighbours, hence $Z_c^{\rm{lat}} = \langle Z_c\rangle = 4$. In contrast, in the AB$_2$ lattice shown in main text Fig.~4a, a third of the particles have six bonded neighbours and two thirds have three neighbours, thus in this system  $Z_c^{\rm{lat}} = \langle Z_c\rangle = \tfrac{6}{3} + \tfrac{2\times3}{3} = 4$. 
	
	In a non-infinite system, the average coordination number will never reach $Z_c^{\rm{lat}}$ as the particles on the boundaries will have less bonds, and thus $\langle Z_c\rangle<Z_c^{\rm{lat}}$. If we assume that the boundaries are on average straight, we can write that the average coordination number at the boundary is $\langle Z_c^{\rm{boundary}}\rangle=Z_c^{\rm{lat}}/2$. Thus the average coordination number can be estimated by:
	\begin{align}
		\langle Z_c\rangle &= \left[(N - N_{\rm{boundary}})Z_c^{\rm{lat}} + \frac{N_{\rm{boundary}}Z_c^{\rm{lat}}}{2}\right]/N\\
		&= Z_c^{\rm{lat}}\left[1 - \frac{N_{\rm{boundary}}}{2N}\right]
	\end{align}
	Where $N_{\rm{boundary}}$ is the number of particles on the boundary and $N$ is the total number of particles. We define $\gamma = N_{\rm{boundary}}/N$ to arrive at Eq.1 in the main text. 
	
	The boundary to bulk ratio $\gamma$ depends on the shape and size of the clusters. Obviously for a cluster that is a simple chain of $N$ particles, $\gamma = 1$ as all particles are on the boundary. For a contiguous, convex, isotropic cluster of size $N$ we can estimate that the area of the cluster scales with $N$ and the boundary scales with $N^{1/2}$, thus in the high $N$ limit $\gamma$ would scale with $N^{-1/2}$. Thus smaller clusters give a higher value of $\gamma$ and would give smaller estimates of $\langle Z_c\rangle$.
	
	\newpage
	\section{Supplemental simulation results}
	
	\subsection{Response of coordination number to size and number imbalance}
	
	We measure the average coordination number, $\langle Z_c\rangle$, in a simulation of $100$ particles after $10^7$ time steps, shown in Fig.~\ref{fig:coord}. We see that $\langle Z_c\rangle$ is maximized when the size and number imbalance is small, which is where the system evolves towards a hexagonal arrangement. $\langle Z_c\rangle$ is minimized when $N_a/N_b\approx R_a/R_b < 1$. In this regime, the system targets a square lattice, thus $Z_c^{\rm{lat}} = 4$, furthermore the clusters are smaller so $\gamma$ is increased. 
	
	\begin{figure}[h!]
		\centering
		\includegraphics [width=0.5\textwidth]{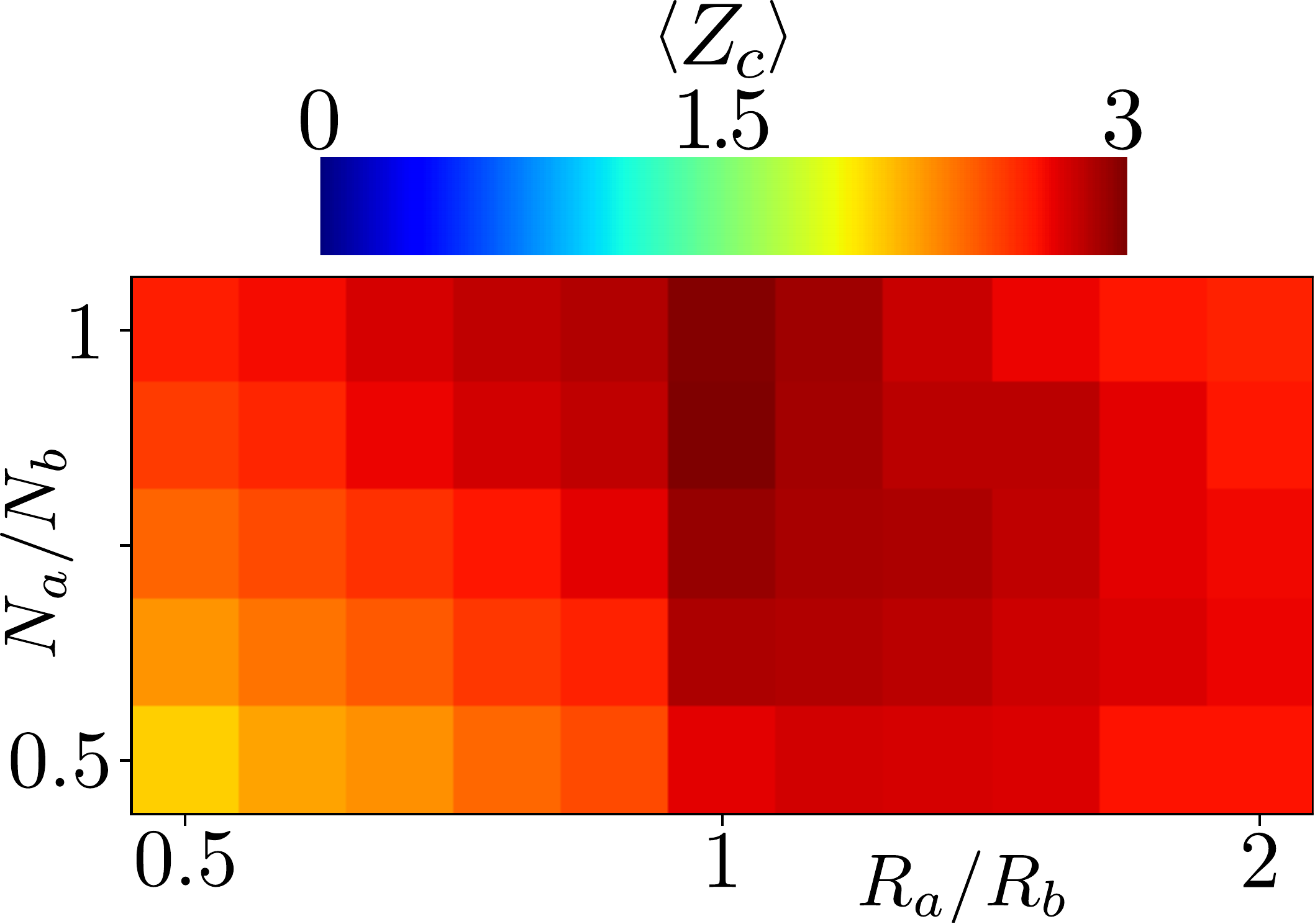} 
		\caption{Average coordination number $\langle Z_c\rangle$ for a simulation of $100$ circular particles with varying size $R_a/R_b$ and number ratio $N_a/N_b$. }
		\label{fig:coord}
	\end{figure}
	
	\subsection{Cluster growth and coordination number in simulations}
	
	To make the simulated results presented in Fig.2c, 3c and 5c comparable we match the rate of growth of the average coordination number, see Fig.~\ref{fig:coord}a. This point of comparison is chosen to account for the irregular shapes of the clusters featuring square particles. Clusters featuring square particles are typically long and thin, owing to their reduced ability to reconfigure to a more compact configuration. This means that for the same cluster size, $\gamma$ and therefore $\langle Z_c\rangle$ are very low, making  the estimate of $\psi_4$ and $\psi_6$ less accurate. Indeed, for a very similar $\langle Z_c\rangle$ (shown in Fig.~\ref{fig:coord} a) the average cluster size $\langle N_c\rangle$ (shown in Fig.~\ref{fig:coord} b) is much larger for cube-sphere systems than for systems consisting of two circular particles. 
	
	\begin{figure}[h!]
		\centering
		\includegraphics [width=0.75\textwidth]{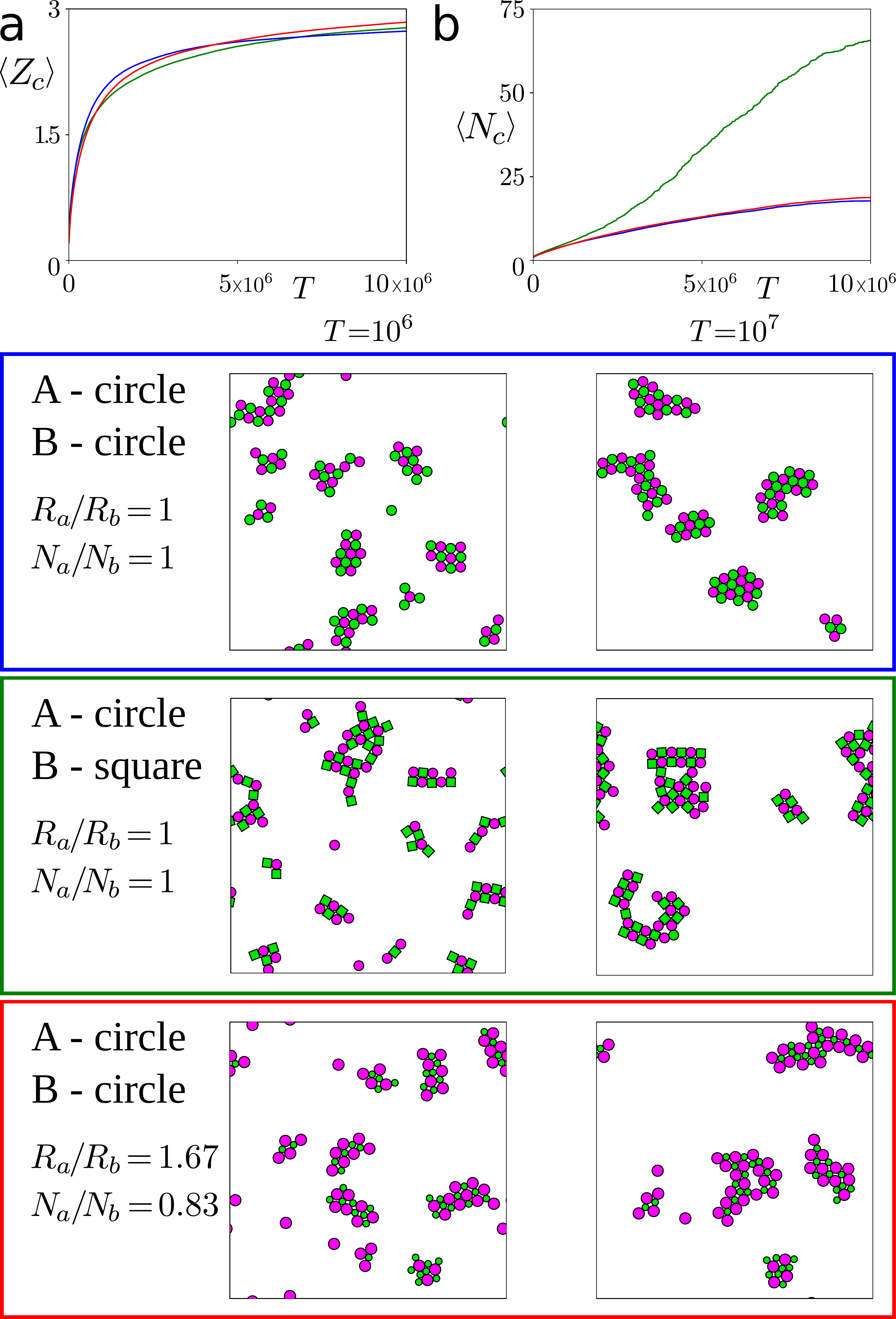} 
		\caption{(a) Average coordination number $\langle Z_c\rangle$ and (b) cluster size $\langle N_c\rangle$ for simulations used in main text Fig.2c (blue), 3c (red), and 5c (green). Below, snapshots in the early (left) and late (right) stages of a simulation for each set of parameters.}
		\label{fig:coord}
	\end{figure}

\end{document}